\documentclass[traditabstract]{aa} 
\usepackage{graphicx}
\usepackage{txfonts}
\usepackage{subfigure}
\usepackage{multirow}
\usepackage{natbib}
\usepackage{array}
\usepackage{soul}
\usepackage[normalem]{ulem}

\usepackage{lscape}
\begin{document}
   \title{Discovery of \ion{O}{vii} line emitting gas in elliptical galaxies}


   \author{Ciro Pinto
          \inst{1}
          \and
          Andrew~C. Fabian \inst{1}
          \and
          Norbert Werner \inst{2,3}
          \and
          Peter Kosec \inst{1,2,3}
          \and
          Jussi Ahoranta \inst{4}
          \and
          Jelle de Plaa \inst{5}
          \and\\
          Jelle~S. Kaastra\inst{5}
          \and
          Jeremy~S. Sanders \inst{6}
          \and
          Yu-Ying Zhang \inst{7}
          \and
          Alexis Finoguenov \inst{4}
          }

   \institute{Institute of Astronomy, Madingley Road, CB3 0HA Cambridge, United Kingdom, \email{cpinto@ast.cam.ac.uk}.
         \and
             Kavli Institute for Particle Astrophysics and Cosmology, Stanford University, 452 Lomita Mall, Stanford, CA 94305-4085, USA
         \and
             Department of Physics, Stanford University, 382 Via Pueblo Mall, Stanford, CA 94305-4060, USA
         \and
             Department of Physics, University of Helsinki, FI-00014 Helsinki, Finland
         \and    
             SRON Netherlands Institute for Space Research, Sorbonnelaan 2, 3584 CA Utrecht, The Netherlands.
         \and
             Max-Planck-Institut fur extraterrestrische Physik, Giessenbachstrasse 1, D-85748 Garching, Germany
         \and
             Argelander-Institut f\"ur Astronomie, Universit\"at Bonn, Auf dem H\"ugel 71, 53121 Bonn, Germany. 
             }
   \date{Received November 4, 2014 / Accepted November 11, 2014}

  \abstract
   {In the cores of ellipticals, clusters, and groups of galaxies, the gas has a cooling time shorter than 1\,Gyr.
    It is possible to probe cooling flows through the detection of \ion{Fe}{xvii} and \ion{O}{vii}
    emission lines, but so far \ion{O}{vii} has not been detected in any individual object.
    The Reflection Grating Spectrometers (RGS) aboard XMM-\textit{Newton} are currently the only instruments
    able to detect \ion{O}{vii} in extended objects such as elliptical galaxies and galaxy clusters. 
    We searched for evidence of \ion{O}{vii} through all the archival RGS observations of galaxy clusters, 
    groups of galaxies, and elliptical galaxies
    focusing on those with core temperatures below 1\,keV.
    We have discovered \ion{O}{vii} resonance (21.6\,{\AA}) and forbidden (22.1\,{\AA}) lines 
    for the first time in the spectra of individual objects.
    \ion{O}{vii} was detected at a level higher than three sigma in six elliptical galaxies:
    M\,84, M\,86, M\,89, NGC\,1316, NGC\,4636, and NGC\,5846.
    M\,84, M\,86, and M\,89 are members of the Virgo Cluster, the others are central dominant galaxies of groups,
    and most them show evidence of \ion{O}{vi} in UV spectra.
    We detect no significant trend between the \ion{Fe}{xvii} 
    and \ion{O}{vii} resonance-to-forbidden line ratios, possibly because of the limited statistics. 
    The observed line ratios $<{\rm Fe}_{r/f},{\rm O}_{r/f}>=(0.52\pm0.02,0.9\pm0.2)$
    indicate that the spectra of all these ellipticals are affected by resonance scattering, suggesting low turbulence.
    Deeper exposures will help to understand whether the \ion{Fe}{xvii} 
    and \ion{O}{vii} lines are both produced by the same cooling gas
    or by multiphase gas.
    Our \ion{O}{vii} luminosities correspond to $0.2-2\,M_{\odot}$\,yr$^{-1}$,
    which agree with the predictions for ellipticals. 
    Such weak cooling rates would not be detected in clusters 
    because their spectra are dominated by the emission of hotter gas, 
    and owing to their greater distance, the expected \ion{O}{vii} line flux would be undetectable.}


   \maketitle
%

\section{Introduction}
\label{sec:introduction}

\vspace{-0.1cm}

Most of the baryonic mass in clusters of galaxies is found in the form of hot $10^{6-8}$\,K gas, 
known as intracluster medium (ICM). The density of this gas strongly increases in the cores 
of the clusters where the radiative cooling time is less than 1\,Gyr. In the absence of heating,
this would imply hundreds of solar masses per year cooling below $10^{6}$\,K \citep{Fabian1994}. 
At these temperatures, the gas is expected to produce prominent emission lines 
from \ion{O}{vi} in UV, as well as \ion{O}{vii} and \ion{Fe}{xvii} in X-rays.
\ion{O}{vi} and \ion{O}{vii} emissivities peak at $5\times10^5$\,K and $2\times10^6$\,K, respectively.

Weak cooling flows have been detected through the \ion{O}{vi} UV lines by 
\cite{Bregman2005,Bregman2006} at the levels of $30\,M_{\odot}$\,yr$^{-1}$
or even lower \citep{McDonald2014} compared to the predicted $10^2M_{\odot}$\,yr$^{-1}$. 
\ion{Fe}{xvii} emission lines have been discovered, but no significant detections of \ion{O}{vii} 
in any galaxy, cluster, or group of galaxies have been reported so far. Therefore there is a strong deficit of gas below 0.5\,keV and
much less cool gas than expected from cooling-flow models (see, e.g., \citealt{Peterson2003}). 
Recently, \cite{Sanders2011_OVII} have stacked 4.6\,Ms of high-spectral-resolution 
XMM-\textit{Newton} Reflection Grating Spectrometer (RGS) spectra from galaxy clusters, groups of galaxies,
and elliptical galaxies. For those objects with a core temperature below 1\,keV, they detected \ion{O}{vii} emission 
for the first time, but they pointed out that the \ion{O}{vii}/\ion{Fe}{xvii} line flux ratio
is at least one fourth of what is expected for isobaric radiative cooling. Therefore, there is either
some heating process preventing cooling below 0.5\,keV or absorbing material around the coolest 
X-ray emitting gas, or non-radiative cooling takes place
(see, e.g., \citealt{Sanders2011_OVII} and \citealt{Werner2013}).

The RGS instruments aboard XMM-\textit{Newton} are currently the only high-resolution 
X-ray spectrometers able to detect and resolve the \ion{O}{vii} and \ion{Fe}{xvii} emission lines.
We searched for evidence of \ion{O}{vii} throughout all the archival XMM-\textit{Newton} grating 
observations as published by 2014 August.
We particularly focused on those objects that exhibit low-temperature ($<1$\,keV) emitting gas
(see, e.g., \citealt{Su2013}).
We detected \ion{O}{vii} in six objects at a level $>3\sigma$,
but find some evidence of \ion{O}{vii} in a total of nine elliptical galaxies.
We benefited from excellent new data, in particular for NGC\,5846, 
provided by the CHEERS project (see de Plaa et al. in prep and Pinto et al. submitted).

This letter summarizes our results.
In Sect.~\ref{sec:data} we present the data reduction.
The spectral modeling is reported in Sect.~\ref{sec:modeling}. 
We discuss the results in Sect.~\ref{sec:discussion} and 
give our conclusions in Sect.~\ref{sec:conclusion}.

\vspace{-0.5cm}

\section{The data}
\label{sec:data}

\vspace{-0.1cm}

The observations used in this paper are listed in Table~\ref{table:log}. 
The XMM-\textit{Newton} satellite is equipped with two types of X-ray detectors: 
the CCD type European Photon Imaging Cameras (EPIC) and the Reflection Grating Spectrometers (RGS).
The RGS spectrometers are slitless, and the spectral lines are broadened by the source extent.
We correct for spatial broadening through the use of EPIC/MOS\,1 surface brightness profiles.

All the observations have been reduced with the 
XMM-\textit{Newton} Science Analysis System (SAS) v13.5.0
using the latest calibration files.
We correct for contamination from soft-proton flares following the XMM-SAS standard procedures.

For each source, we extracted the first-order RGS spectra in two cross-dispersion regions of 3.4' and 0.8' widths 
centered on the emission peak. These regions cover
the bulk of the galaxy emission and the cores of the galaxies, respectively.
We subtracted the model background spectrum, which is created by the standard RGS pipeline 
and is a template background file based on the count rate in CCD\,9.
For each source, we stacked the RGS spectra of all the observations with the \textit{rgscombine} task
and converted them to SPEX format through the SPEX task \textit{trafo}.
We extracted the MOS\,1 images in the $8-27$\,{\AA} wavelength band and obtained
surface brightness profiles to model the RGS line spatial broadening
(see e.g. \citealt{dePlaa2012}).

Our analysis focuses on the $8-27$\,{\AA} first-order RGS spectra.
The spectra are binned by a factor of 3, which provides the optimal 
spectral bin size of about 1/3\,FWHM.
We perform the spectral analysis with SPEX\footnote{www.sron.nl/spex}
version 2.03.03 and scale elemental abundances to the proto-solar abundances 
of \citet{Lodders09}. 
We use the updated ionization balance calculations of \citet{Bryans2009}.
C-statistics and $1\,\sigma$ errors are adopted.

\begin{table}
\caption{XMM-\textit{Newton}/RGS observations used in this paper.}  
\vspace{-0.2cm}
\label{table:log}      
\renewcommand{\arraystretch}{1.1}
 \small\addtolength{\tabcolsep}{-3pt}
 
\scalebox{1}{%
\begin{tabular}{c c c c c c c}     
\hline\hline                                                                                                                                                                             
Source                              & ID $^{(a)}$ & t\,(ks)$^{(b)}$            & $z$                       & \ion{O}{vi}          & \ion{O}{vii}$^{tot}$      & \ion{O}{vii}$^{sim}$     \\  
\hline                                                                                                                                                                                                               
\multirow{1}{*}{\object{IC 1459}}   & 0135980201  &  26                        & \multirow{1}{*}{{0.006}}  & Y                    & $2.5$                     & $1.5$                    \\
\hline                                                                                                                                                                                   
\multirow{1}{*}{\object{M 84}}      & 0673310101  &  64                        & \multirow{1}{*}{{0.0034}} & Y                    & $3.7$                     & $3.4$                    \\
\hline                                                                                                                                                                                   
\multirow{1}{*}{\object{M 86}}      & 0108260201  &  91                        & \multirow{1}{*}{{-0.0009}}& P                    & $4.1$                     & $6.5$                    \\
\hline                                                                                                                                                                                   
\multirow{1}{*}{\object{M 89}}      & 0141570101  &  29                        & \multirow{1}{*}{{0.001}}  & P                    & $3.1$                     & $3.3$                    \\
\hline                                                                                                     
\multirow{2}{*}{\object{NGC 1316}}  & 0302780101  & \multirow{2}{*}{{166}}     & \multirow{2}{*}{{0.0059}} & \multirow{2}{*}{{Y}} & \multirow{2}{*}{{$6.3$}}  & \multirow{2}{*}{{$6.7$}} \\
                                    & 0502070201  &                            &                           &                      &                           &                          \\
\hline                                                                                                                                                                                   
\multirow{1}{*}{\object{NGC\,1332}} & 0135980201  &  64                        & \multirow{1}{*}{{0.0052}} & UL                   & $2.1$                     & $1.6$                    \\
\hline                                                                                                                                                                                   
\multirow{1}{*}{\object{NGC\,1404}} & 0304940101  &  29                        & \multirow{1}{*}{{0.0065}} & UL                   & $2.0$                     & $2.0$                    \\
\hline                                                                         
\multirow{2}{*}{\object{NGC\,4636}} & 0111190101  & \multirow{2}{*}{{102}}     & \multirow{2}{*}{{0.0037}} & \multirow{2}{*}{{Y}} & \multirow{2}{*}{{$5.0$}}  & \multirow{2}{*}{{$6.0$}} \\
                                    & 201/501/701 &                            &                           &                      &                           &                          \\
\hline   
\multirow{2}{*}{\object{NGC\,5846}} & 0021540101/501  & \multirow{2}{*}{{195}} & \multirow{2}{*}{{0.0061}} & \multirow{2}{*}{{Y}} & \multirow{2}{*}{{$3.6$}}  & \multirow{2}{*}{{$3.7$}} \\
                                    & 0723800101/201  &                        &                           &                      &                           &                          \\
\hline   
\end{tabular}}

$^{(a, \, b)}$ Exposure ID and total RGS clean time.
\ion{O}{vi} detections are taken from Bregman et al. (2005): Y=yes, P=possible, UL=upper limits.
\ion{O}{vii} combined significance in units of $\sigma$: 21.6\,{\AA} and 22.1\,{\AA} significance added in quadrature ($tot$)
             and simultaneous fit ($sim$) with $f_{21.6/22.1}\equiv1.3$  
             as predicted by our thermal model (see Table\,\ref{table:fits} and Sect.\,\ref{sec:significance}).\\
\vspace{-0.75cm}

\end{table}                             

\begin{table*}
\caption{XMM-\textit{Newton}/RGS cooling flow modeling and \ion{O}{vii} local fits.}  
\label{table:fits}      
\vspace{-0.2cm}
\renewcommand{\arraystretch}{1.2}
 \small\addtolength{\tabcolsep}{-2pt}
 
\scalebox{1}{%
\begin{tabular}{c| c c c c c| c c |c c c c}     
\hline\hline            
                                    & \multicolumn{5}{c|}{0.8' region}                                                             & \multicolumn{2}{c|}{3.4' region}      & \multicolumn{2}{c}{highest $\sigma_{r\,f}$} & \multicolumn{2}{c}{UV / FUSE}     \\   
\hline                                                                                                                                                                                                                                  
Source                              & $M_{\odot}$yr$^{-1}$ & T\,(keV)      & O/Fe          & \ion{O}{vii}$^r$  & \ion{O}{vii}$^f$ & \ion{O}{vii}$^r$  & \ion{O}{vii}$^f$ & \ion{O}{vii}$^r$  & \ion{O}{vii}$^f$ & \ion{O}{vi}           & $M_{\odot}$yr$^{-1}$\\  
\hline                                                                                                                                                                                                                         
\multirow{1}{*}{\object{IC 1459}}   & $0.22\pm0.04$        & $0.79\pm0.09$ & $1.50\pm0.30$ & $0.04\pm0.04$     & $-$              & $0.18\pm0.07$*    & $-$*             & $2.5$             & $-$              & Y                     & 1.55 \\
\multirow{1}{*}{\object{M 84}}      & $0.48\pm0.02$        & $1.10\pm0.04$ & $0.79\pm0.06$ & $0.11\pm0.03$*    & $0.03\pm0.02$*   & $0.03\pm0.03$     & $-$              & $3.5$             & $1.2$            & Y                     & 0.32 \\
\multirow{1}{*}{\object{M 86}}      & $0.55\pm0.02$        & $1.35\pm0.06$ & $0.89\pm0.07$ & $0.29\pm0.08$     & $-$              & $0.50\pm0.16$*    & $0.48\pm0.19$*   & $3.2$             & $2.6$            & P                     & 0.26 \\
\multirow{1}{*}{\object{M 89}}      & $0.30\pm0.03$        & $0.83\pm0.06$ & $0.94\pm0.13$ & $-$               & $0.09\pm0.04$    & $0.15\pm0.09$*    & $0.24\pm0.09$*   & $1.7$             & $2.6$            & P                     & 0.24 \\
\multirow{1}{*}{\object{NGC 1316}}  & $0.30\pm0.01$        & $1.07\pm0.04$ & $0.90\pm0.06$ & $0.12\pm0.02$*    & $0.07\pm0.02$*   & $0.13\pm0.03$     & $0.04\pm0.03$    & $5.2$             & $3.5$            & \multirow{1}{*}{{Y}}  & 0.78 \\
\multirow{1}{*}{\object{NGC\,1332}} & $0.24\pm0.02$        & $0.83\pm0.06$ & $0.60\pm0.12$ & $0.03\pm0.02$*    & $0.03\pm0.02$*   & $-$               & $-$              & $1.5$             & $1.5$            & UL                    & $-$  \\
\multirow{1}{*}{\object{NGC\,1404}} & $1.19\pm0.04$        & $0.97\pm0.03$ & $0.56\pm0.05$ & $0.07\pm0.06$*    & $0.10\pm0.06$*   & $-$               & $-$              & $1.1$             & $1.7$            & UL                    & 0.19 \\
\multirow{1}{*}{\object{NGC\,4636}} & $1.14\pm0.02$        & $1.06\pm0.02$ & $0.57\pm0.02$ & $0.09\pm0.04$     & $0.14\pm0.04$    & $0.23\pm0.07$*    & $0.28\pm0.07$*   & $3.2$             & $3.9$            & \multirow{1}{*}{{Y}}  & 0.27 \\
\multirow{1}{*}{\object{NGC\,5846}} & $1.77\pm0.02$        & $1.13\pm0.02$ & $0.62\pm0.03$ & $0.07\pm0.02$*    & $0.03\pm0.02$*   & $0.08\pm0.03$     & $0.03\pm0.02$    & $3.3$             & $1.5$            & \multirow{1}{*}{{Y}}  & 0.87 \\
\hline   
\end{tabular}}

Notes: Cooling rates, temperatures, and O/Fe abundance ratios for the 8--27\,{\AA} RGS cooling flow model. 
The flux and significance of each line were measured with accurate local fits (see Sect.\,\ref{sec:significance}). 
Line fluxes are in photons\,m$\,^{-2}$\,s$\,^{-1}$. 
Starred fluxes (*) are those maximizing both \ion{O}{vii} $(r,f)$ detections.
UV \ion{O}{vi} detections and cooling rates (with typical $\gtrsim$30\% uncertainties) 
are taken from Bregman et al. (2005, see also Table\,\ref{table:log}).\\
\end{table*}                             

\vspace{-0.4cm}

\section{Spectral modeling}
\label{sec:modeling}

We have modeled the RGS spectra with a cooling flow model (\textit{cflow}),
adopting a 0.1\,keV minimum temperature. The \textit{cflow} is corrected for the redshift
and the Galactic absorption. 
The column densities are estimated through the tool of \citet{Willingale2013}
to include contribution to absorption from both atomic and molecular hydrogen. 
The spectral model was broadened by the source extent through the \textit{lpro} component that
receives the MOS\,1 spatial profile as input.
For some sources we also added an absorbed, redshifted powerlaw 
to account for emission from the central point-like source.
Free parameters in the fits are the cooling rate, the upper temperature, 
and the O/Fe, N/Fe, Ne/Fe, and Mg/Fe abundance
ratios (Fe and other elements are assumed solar).
We did not explicitly model the cosmic X-ray background in RGS because any diffuse emission feature 
would be smeared out into a broad continuum-like component. 

The best-fit cooling rates, upper temperatures, and O/Fe abundance ratios 
are shown in Table\,\ref{table:fits}.
On average, the fits provide C-stat/dof=540/400. We computed an equivalent $\chi^2_r\sim1.35$.
Contrary to hotter groups and clusters of galaxies (see, e.g., \citealt{Peterson2003}),
the \textit{cflow} model describes the spectra of these cold systems very well.
Their cooling rates are between $0.2$ and $2\,M_{\odot}$yr$^{-1}$,
which are typically predicted for elliptical galaxies (see, e.g., \citealt{Sarazin1989}).
Our values are generally larger than those measured by \citet{Bregman2005}, except for IC\,1459 and NGC\,1316,
most likely because FUSE has a smaller (30'') aperture than our 50'' and 200'' regions.
The disagreement between our \ion{Fe}{xvii}--\ion{O}{vii} and their \ion{O}{vi} cooling rates
may be due to the different assumptions made; for instance, they adopted solar abundances
and the luminosities of one ion, \ion{O}{vi}.

\vspace{-0.4cm}

\subsection{{\ion{O}{vii}} detection level}
\label{sec:significance}

To measure the significance of the {\ion{O}{vii}} 
emission lines, we have removed the {\ion{O}{vii}} ion from the model
and fitted two delta lines fixed at 21.6\,{\AA} and 22.1\,{\AA}, which reproduce the {\ion{O}{vii}} 
resonance and forbidden lines, respectively.
The intercombination line at 21.8\,{\AA} generally appears to be insignificant 
and blends with the resonance line.
These lines are corrected by the redshift, the Galactic absorption,
and the spatial broadening, as in the \textit{cflow} model.
We tabulate the fluxes of these lines as measured in the 3.4' and 0.8' regions
in Table\,\ref{table:fits}. We quote the detection confidence levels
for the regions with the highest significance for both {\ion{O}{vii}} lines.
We detected the {\ion{O}{vii}} 21.6\,{\AA} resonance line at a $\gtrsim3\sigma$ level in five objects.
The significance of the two lines was also combined by adding them in quadrature (see \ion{O}{vii}$^{tot}$
in Table\,\ref{table:log}).
In order to have more robust results, we have fitted again the lines by fixing their flux ratio
to $(r/f)\equiv1.3$ as predicted by the thermal model (see \ion{O}{vii}$^{sim}$
in Table\,\ref{table:log}). The combination of the lines generally strengthen
the significance and in particular provides $\gtrsim3\sigma$ for M\,89.

\begin{figure*}
\vspace{-0.8cm}
  \begin{center}
      \subfigure{ 
      \includegraphics[bb=30 10 523 740, width=0.675\textwidth, angle=-90]{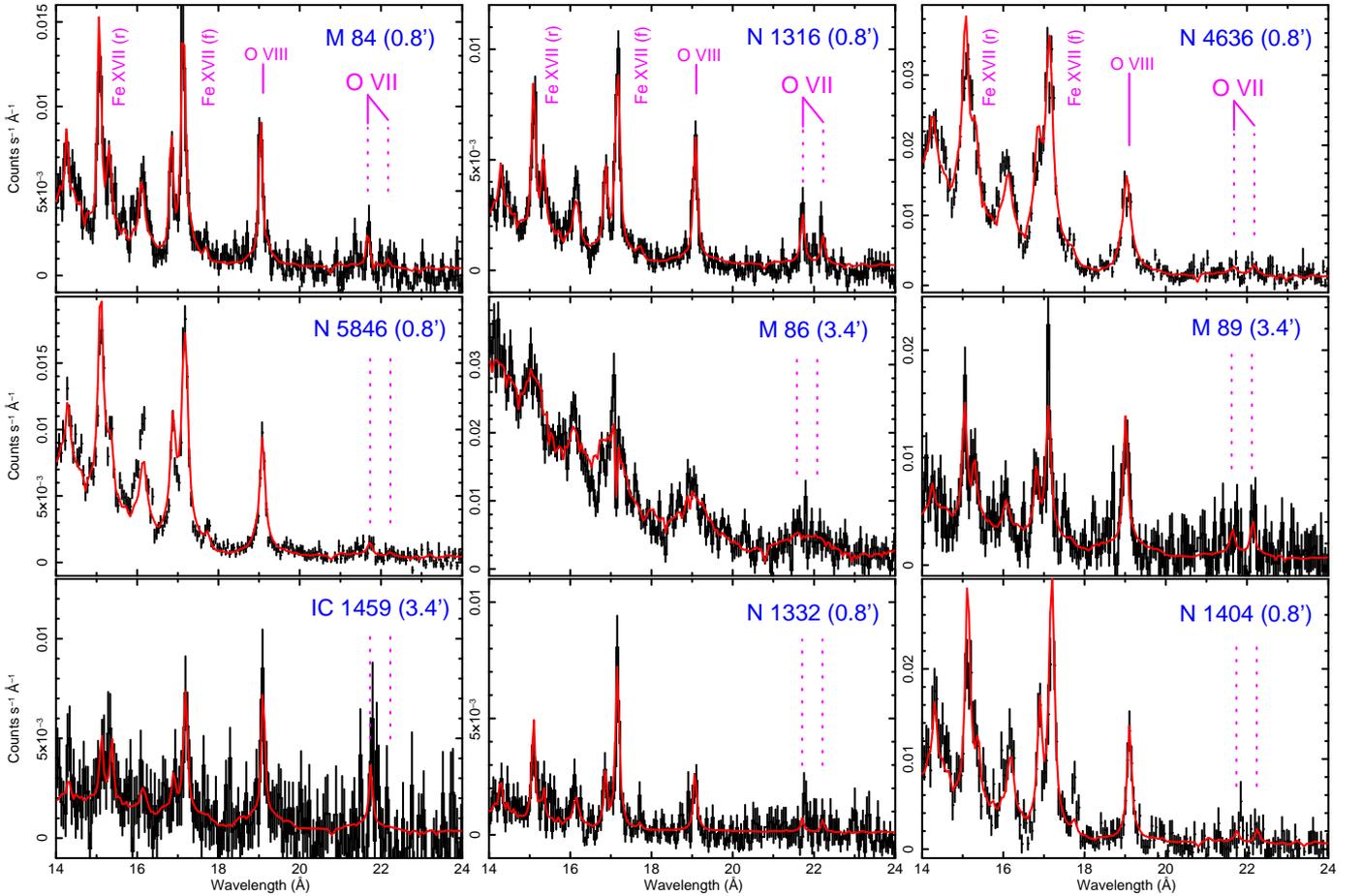}}
\vspace{-0.4cm}
      \caption{XMM-\textit{Newton}/RGS spectra with cooling flow models.
               \ion{O}{vii} and \ion{Fe}{xvii} are fitted separately.
               The \ion{O}{vii} significance increases upward and 
               the 22.1\,{\AA}/21.6\,{\AA} line ratio increases rightward.
               The extraction regions and the rest-frame wavelengths of the 
               \ion{O}{vii} lines are also indicated.}
          \label{fig:fits}
  \end{center}
\vspace{-0.8cm}
\end{figure*}

We did the same exercise for the \ion{Fe}{xvii} emission lines 
by removing the \ion{Fe}{xvii} ion from the model and fitting four delta lines fixed at
15.01\,{\AA}, 15.26\,{\AA}, 16.78\,{\AA}, and 17.08\,{\AA},
which are the main \ion{Fe}{xvii} transitions.
We did not tabulate the significance of the \ion{Fe}{xvii} lines because
they are much stronger than the {\ion{O}{vii}} lines with 
errors on average smaller by a factor $\sim5$ (see Sect.\,\ref{sec:discussion}).
In Fig.\,\ref{fig:fits} we show the final spectral fits 
with the specific modeling of the oxygen and the iron lines.
The plots focus on the 14--24\,{\AA} wavelength range, which includes the relevant \ion{Fe}{xvii}
and {\ion{O}{vii}} lines. The remaining emission lines are reproduced 
by the cooling-flow model.

\vspace{-0.4cm}

\subsection{Further tests}

The continuum modeling may affect the \ion{O}{vii} line fluxes. 
As for the test, we locally refit the 18--23\,{\AA}
3.4' NGC\,4636 and 5846 spectra.
We still obtain $>2.3\sigma$ detections for the 21.6\,{\AA} line in NGC\,5846
and for both the lines in NGC\,4636.
NGC\,1316--4636--5846 have multiple observations. We confirm the detection
of {\ion{O}{vii}} in the RGS\,1 spectra of the two individual observations of NGC\,1316
and in all the observations of NGC\,4636 and NGC\,5846
with exposure times above 40\,ks. 

An accurate background subtraction is crucial in this analysis.
Our model background spectra are estimated from blank field observations and scaled by the 
count rate in CCD\,9 (5--8\,{\AA}), where hardly any emission from each source is expected
(see Sect.\,\ref{sec:data}). 
Galaxies, groups and clusters of galaxies may 
still have significant emission in the 5--8\,{\AA} range.
Our background subtraction may have been more severe than the necessary one, 
but this implies that our detection levels are conservative.

We also tested a different background 
for the 0.8' and 3.4' NGC\,1316--4636--5846 spectra
by selecting the standard region outside the 98\% of the source counts. 
This background subtraction decreases
the statistics in the 20--24\,{\AA} range because 
it includes part of the source emission,
but we still detect the 21.6\,{\AA} line in each  
well-exposed RGS\,1 spectrum of NGC\,1316 and 5846.
For NGC\,4636 we only detect the 22.1\,{\AA} line.

\vspace{-0.7cm}

\section{Discussion}
\label{sec:discussion}

\vspace{-0.1cm}

The previous detection of {\ion{O}{vii}} in RGS stacked spectra of different sources
and the FUSE/UV results on {\ion{O}{vi}} in E-type galaxies motivated us to search for {\ion{O}{vii}} 
in the high-resolution X-ray spectra of cool ($\lesssim1$\,keV) systems.
We searched throughout all the RGS archive of clusters, groups,
and elliptical galaxies.
In this paper we have shown the results for those objects that show
some evidence of {\ion{O}{vii}}. This is the first time that {\ion{O}{vii}} 
is significantly detected in a spectrum of an E-Type galaxy.

We detect the {\ion{O}{vii}} $(r)$ 21.6\,{\AA} line at a $\gtrsim3\sigma$ level in five objects:
M\,84, M\,86, NGC\,1316, NGC\,4636, and NGC\,5846. 
IC\,1459, M\,89, NGC\,1332, and NGC\,1404 show evidence of {\ion{O}{vii}} but 
with lower significance. This is not surprising since they are typically
fainter than the former five sources (see Fig.\,\ref{fig:fits}),
or their observations have shorter exposure times (see Table\,\ref{table:log}).
Deeper exposures should increase the significance of the detection
and possibly unveil {\ion{O}{vii}} in those sources whose exposures
were too short to provide useful constraints. 
The combined  21.6\,{\AA}--22.1\,{\AA} significance is $\gtrsim3\sigma$
for M\,89.

For some galaxies the {\ion{O}{vii}} lines are only detected in one region.
The 21.6\,{\AA} line is not significantly detected in the 0.8' spectra of 
IC\,1459, M\,89, and NGC\,1404, most likely because
of the lower statistics in the narrow core.
For the other five objects the (cool) 0.8' core provides a higher significance
presumably because the hotter lines like {\ion{O}{viii}} have less of an impact on the statistics,
and the counts of the {\ion{O}{vii}} lines are still high.
M\,86 has a broader core than the eight other sources,
which makes the lines broader and the detection more difficult.
This is clearly seen in the 3.4' spectrum where the lines are highly broadened,
and it is not possible to resolve the 21.6\,{\AA} and 22.1\,{\AA} lines
(see Fig.\,\ref{fig:fits} and Table\,\ref{table:fits}).
We find a $>2\sigma$ detection for all the sources showing {\ion{O}{vi}} 
(\citealt{Bregman2005}, see also Table\,\ref{table:fits}).

\vspace{-0.4cm}

\subsection{Evidence of resonant scattering}

For some objects, in particular NGC\,4636, the {\ion{O}{vii}} 22.1\,{\AA} forbidden line appears to be stronger
than the resonance line, though they would agree within the errors. 
In the absence of chaotic motions, the gas is optically thick in the resonant lines
due to resonant scattering, while the forbidden lines remain optically thin.
In the cores of NGC\,4636 and 1404, the \ion{Fe}{xvii} 15.0\,{\AA} 
line is strongly suppressed by resonant scattering as found by \citet{Werner2009}. 
Their \ion{Fe}{xvii} 15.0\,{\AA} / 17.0\,{\AA} line ratios
agree very well with our estimates.

\vspace{-0.4cm}

\subsection{Origin of the {\ion{O}{vii}} lines}

We believe that the {\ion{O}{vii}} lines are produced by diffuse gas in the galaxies and their groups 
for the following two reasons: 
1) the lines are at their expected, rest-frame wavelengths, and they are too narrow to belong to either the local hot bubble
   or the Galactic gas; 
2) they are broadened by the source extent exactly as the other lines (see Fig.\,\ref{fig:fits}),
   which means that the central point-like AGN would only make a small contribution.

\begin{figure}
  \begin{center}
      \subfigure{ 
      \includegraphics[bb=70 75 520 670, width=6.5cm, angle=+90]{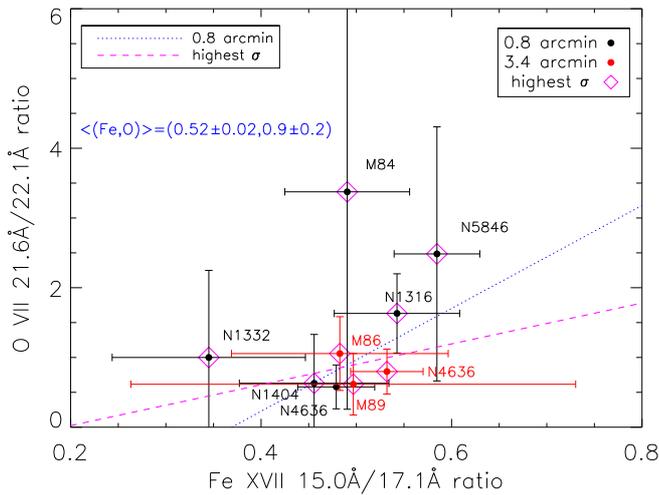}}
      \caption{\ion{O}{vii} VS \ion{Fe}{xvii} resonance-to-forbidden line ratios. 
               The least-squares for the 0.8' region and the combined 0.8'$-$3.4' 
               highest-$\sigma$ points are shown (see also Table\,\ref{table:fits}). 
               The average (Fe,O) values are quoted.}
          \label{fig:ovii_ratios}
  \end{center}
\vspace{-0.75cm}
\end{figure}

It is more difficult to understand whether the Fe and O lines are produced by the same gas.
We find that a single cooling flow model is able to fit the lines very well for all the objects.

We calculated the resonance-to-forbidden $(r)/(f)$ line ratios of the {\ion{O}{vii}} and \ion{Fe}{xvii} lines as
measured in the 0.8' core and compared them in Fig.\,\ref{fig:ovii_ratios}.
If resonant scattering is occurring similarly for the 21.6\,{\AA} and the 15.0\,{\AA} lines,
then we should find some correlation. 
The limited statistics of the {\ion{O}{vii}} $(r)/(f)$ line ratio
does not allow us to put strong constraints.
However, we find a weak correlation between the oxygen and iron line ratios,
but their values agree with an average 
$<{\rm Fe}_{r/f},{\rm O}_{r/f}>=(0.52\pm0.02,0.9\pm0.2)$.
Deeper exposures will help to search for any clear trend in O--Fe $(r)/(f)$ line ratios.

Interestingly, our optically thin plasma models predict ${\rm O}_{r/f}>1.1$ and ${\rm Fe}_{r/f}>0.7$,
depending slightly on the temperature. Our low values for ${\rm Fe}_{r/f}$
indicate that all these ellipticals are affected by resonant scattering.
This would suggest low turbulence in these objects as found by Pinto et al. (submitted).

It is not possible to detect $\lesssim1\,M_{\odot}$yr$^{-1}$ in cluster spectra
because the corresponding emission lines would be hiding underneath the hot gas continuum.
The $50-100\,M_{\odot}$yr$^{-1}$ predicted rates are in principle detectable in RGS spectra of brightest cluster galaxies, 
but they have not been found. Dissipation of turbulence has recently been proposed 
as a mechanism that prevents cooling in clusters (see, e.g., \citealt{Zhuravleva2014}).

\vspace{-0.4cm}

\section{Conclusion}
\label{sec:conclusion}

Cool core clusters and groups of galaxies and elliptical galaxies are thought 
to host gas with cooling times shorter than the \textit{Hubble} time.
Weak cooling flows have been found in the UV/FUSE \ion{O}{vi} spectra of E-type galaxies.
In the X-ray energy band, the \ion{Fe}{xvii} and \ion{O}{vii}
emission lines are good tools for probing cooling gas. 
\ion{O}{vii} has not been detected before in any individual object.
We searched for \ion{O}{vii} through all the XMM-\textit{Newton}/RGS archive 
of galaxy clusters, groups of galaxies, and ellipticals.

We discovered \ion{O}{vii} resonance (21.6\,{\AA}) and forbidden (22.1\,{\AA}) lines
in a small sample of elliptical galaxies that are all contained in galaxy groups. 
A level higher than $3\sigma$ was measured in six galaxies: 
M\,84, M\,86, M\,89, NGC\,1316, NGC\,4636, and NGC\,5846.
The \ion{O}{vii} emission reveals cool diffuse gas in these galaxies and their groups. 
We could not find any significant trend between the \ion{Fe}{xvii} 
and \ion{O}{vii} resonance-to-forbidden line ratios because of the limited statistics.
However, they agree well with $<{\rm Fe}_{r/f},{\rm O}_{r/f}>=(0.52\pm0.02,0.9\pm0.2)$.
These ratios indicate a significant suppression of the \ion{Fe}{xvii} 15.0\,{\AA} resonance line 
and, possibly, of the \ion{O}{vii} 21.6\,{\AA} resonance line as well
by resonant scattering. This implies low gas turbulence.
The weak $\sim1\,M_{\odot}$yr$^{-1}$ cooling flows discovered in these elliptical galaxies
and group of galaxies could not be significantly detected in clusters of galaxies
because of the dominant hot continuum.


\begin{acknowledgements}
This work is based on observations of XMM-\textit{Newton}, an
ESA science mission funded by ESA Member States and the USA (NASA).
YYZ acknowledges the BMWi DLR grant 50~OR~1304. 
\end{acknowledgements}

\vspace{-0.5cm}

\bibliographystyle{aa}
\bibliography{bibliografia} 

\end{document}